\documentclass[aps,prd,preprint,superscriptaddress,showpacs,nofootinbib]{revtex4-1}
\usepackage{graphicx}
\usepackage{amsfonts}
\usepackage{mathrsfs}
\usepackage{epsfig}
\usepackage{slashed}
\usepackage{dcolumn}%

\usepackage{ulem}
\usepackage{color}
\usepackage{caption}
\usepackage{subcaption}
\definecolor{My_red}        {cmyk}{0.00,1.00,1.00,0.20}

\newcommand{\bmat}{\left(\begin{array}}
\newcommand{\emat}{\end{array}\right)}
\newcommand{\beq}{\begin{equation}}
\newcommand{\eeq}{\end{equation}}

\def\bwt{\begin{widetext}}
\def\ewt{\end{widetext}}
\def\be{\begin{equation}}
\def\ee{\end{equation}}
\def\bea{\begin{eqnarray}}
\def\eea{\end{eqnarray}}
\def\bean{\begin{eqnarray*}}
\def\eean{\end{eqnarray*}}
\def\bary{\begin{array}}
\def\eary{\end{array}}
\def\bit{\begin{itemize}}
\def\eit{\end{itemize}}

\def\su5u1{SU(5) \times U(1)}
\def\fsu5u1{SU(5) \times U(1)'}
\def\so10{SO(10)}
\def\sq20{SO(10) \times SO(10)}

\def\bwt{\begin{widetext}}
\def\ewt{\end{widetext}}
\def\be{\begin{equation}}
\def\ee{\end{equation}}
\def\bea{\begin{eqnarray}}
\def\eea{\end{eqnarray}}
\def\bean{\begin{eqnarray*}}
\def\eean{\end{eqnarray*}}
\def\bary{\begin{array}}
\def\eary{\end{array}}
\def\bit{\begin{itemize}}
\def\eit{\end{itemize}}

\def\su5u1{SU(5) \times U(1)}
\def\fsu5u1{SU(5) \times U(1)'}
\def\so10{SO(10)}
\def\sq20{SO(10) \times SO(10)}

\usepackage[centertags]{amsmath}
\usepackage{amssymb}

\def\Zp{Z^{\prime}}

\begin{document}

\title{A new proposal for diphoton resonance  from $E_6$ motivated \\ extra $U(1)$}

\author{Kasinath Das}

\affiliation{Harish-Chandra Research Institute and Regional Centre for
Accelerator-based Particle Physics, Chhatnag Road, Jhusi, Allahabad 211019, India}

\author{Tianjun Li}

\affiliation{Key Laboratory of Theoretical Physics and
Kavli Institute for Theoretical Physics China (KITPC),
Institute of Theoretical Physics, Chinese Academy of Sciences,
Beijing 100190, P. R. China}

\affiliation{
School of Physical Sciences, University of Chinese Academy of Sciences,
Beijing 100049, P. R. China
}

\affiliation{
School of Physical Electronics, University of Electronic Science and Technology of China, 
Chengdu 610054, P. R. China 
}

\author{S. Nandi}

\affiliation{Department of Physics and Oklahoma Center for High Energy Physics,
Oklahoma State University, Stillwater OK 74078, USA}

\author{Santosh Kumar Rai}

\affiliation{Harish-Chandra Research Institute and Regional Centre for
Accelerator-based Particle Physics, Chhatnag Road, Jhusi, Allahabad 211019, India}

\date{\today}

\begin{abstract}

We propose that the diphoton resonance signal indicated by the recent LHC data might also arise from 
the pair productions of vector-like heavy down-type quarks with mass around $750$ GeV and above. 
The vector-like quark decays into an ordinary light quark and a Standard Model singlet scalar. 
The subsequent decay of scalar singlet produces the diphoton excess. Both the vector-like
 quark and singlet scalars appear naturally in the $E_6$, and their masses can be in 
 the TeV scale with a suitable choice of symmetry breaking pattern. The prediction of
such a proposal would be to see an accompanying dijet signal at the same mass 
with similar cross section in the $2\gamma + 2j$ final state and  two dijet resonances at the same mass 
for a $4j$ final state with a cross section, about 100 times larger. Both predictions can be tested 
easily as the luminosity accumulates in the upcoming runs of the LHC.

\end{abstract}


\preprint{HRI-RECAPP-2016-010,OSU-HEP-16-05}

\maketitle

\section{Introduction}

Recent observation of diphoton excess at $750$ GeV  is the only tentative new physics at the 
LHC so far \cite{atlas,CMS:2015dxe}. This observation is reinforced when combined with data of the early $13$ TeV run
with that of events collected at $8$ TeV, though the significance is not large enough to claim any 
discovery \cite{Aaboud:2016tru,CMS:2016owr}. The diphoton  excess cross section is in the range of 
$3 - 10$ femtobarns (fb), with a resonance width which can be small (around a GeV) or 
large (around $45$ GeV). 

A significant number of papers have proposed a solution by assuming the gauge extensions of  the Standard Model (SM)
symmetry \cite{U1extensions,gauge_ext} which results in the required new particle set that can fit the diphoton excess.
A large number of these papers have a model with TeV-scale breaking $U(1)$ where the scalar 
and  exotic fermions carry non-trivial charges under the new $U(1)$ \cite{U1extensions}.
In most of the works the singlet scalar is produced by two gluons through the vector-like quark triangle 
loop (similar to the SM Higgs production via the top quark loop). Then the singlet 
scalar decays into two photons (as well as more dominantly to two gluons) again via the heavy vector-like 
fermion loops. One drawback of this scenario is that the Yukawa coupling of Q to S, $y_{\bar{Q} Q S}$, 
required is non-perturbative to explain the observed level of cross section, or many copies of such 
Q's need to be introduced. In general, this kind of approach explains why the vector-like fermions are
around the TeV scale.

In this work, we propose an alternative production mechanism for this observed diphoton excess 
which can complement the standard production mode mentioned above or have a significant 
event rate by itself. Our mechanism is that the singlet scalar producing this diphoton signal comes from 
the decay of a heavy down-type quark $xd$ which is a  color triplet and an $SU(2)$  singlet with 
an electric charge of $-1/3$. This heavy down-type exotic quark is pair produced dominantly from 
two gluons via strong interactions. There are two things which help us in increasing the 
cross section. The $xd \,\, \overline{xd}$ pair production cross section is much larger than the singlet 
scalar production via the $xd$ loop (unless the Yukawa coupling of the scalar is somewhat 
greater than unity). 
Also, three such $xd$ and $\overline{xd}$ quarks naturally appear in our model based on $E_6$ from 
three fermion families. The singlet scalar is also naturally present. The pattern of symmetry breaking 
that we shall use gives the singlet scalar mass which is close to the $xd$-quark mass. The 
version of $E_6$ model that we use will be discussed in the next section. 

One thing we want to emphasize is that the vector-like quark and singlet scalar, which we are using 
to  explain the diphoton excess, are not postulated in an ad-hoc manner for this purpose, but these 
particles are already present in the $E_6$ grand unified model. All we do is to use the 
appropriate symmetry breaking pattern so that one $U(1)$ in addition to the SM gauge symmetry remain unbroken at 
the TeV scale or even higher. The quantum numbers of all the particles are fixed from the 
$E_6$ symmetry. Because of the underlying $E_6$ symmetry, in addition to explaining the 
diphoton excess, we have several predictions. Along with the diphoton resonant signal, we will also 
have events which have both the diphoton resonance as well as a dijet resonance with similar level 
of cross sections. Also we shall have events with two dijet resonance at the same mass but with 
cross section about 100 times the diphoton resonance. These predictions can be tested as more 
data accumulates at the upcoming 13 TeV LHC runs.

In section \ref{sec:model} below, we discuss our model  and the formalism. In section \ref{sec:analysis}, 
we discuss the phenomenology of our model. This includes the details of how our model can explain 
the diphoton excess, and the other predictions that can be tested in near future. Section \ref{sec:summary}
contains our conclusions and discussions.

\section{ The model and the formalism} \label{sec:model}

Our effective symmetry at the TeV scale is the SM together with an extra $U(1)'$. This extra $U(1)'$ is 
a special subgroup of the $E_6$ Grand Unified Theory (GUT) \cite{gursey, Langacker:2008yv, 
general, PLJW, Erler:2002pr, Kang:2004pp, Kang:2004ix, Kang:2009rd}. We use non-supersymmetric 
$E_6$. $E_6$ is special in the sense that it is anomaly free, as well as has chiral fermions. Its fundamental 
representation reduces under $SO(10)$ as

$$\bf 27 = 16  +10  +1$$

The representation {\bf 16} contains the $15$ SM fermions, as well as a right handed neutrino. It 
decomposes under $SU(5)$ as 

$$\bf 16  =  10  + \bar{5} +1$$

The {\bf 10} representation under $SU(5)$ decomposes as

$$\bf 10 = 5  +\bar{5}  +1  $$

The {\bf 5} contains a color triplet and a $SU(2)_L$ doublet, whereas $\bf \bar{5}$ contains a color anti-triplet and 
another $SU(2)$ doublet, and the $\bf 1$ is a SM singlet. The gauge boson is contained in 
the adjoint $\bf 78$ representation of $E_6$.

The particle content of the $\bf 27$ representation, which contains the SM fermions as well as the  
extra fermions, are shown in the first two columns of Table \ref{E6charge}. For three families of 
fermions, we use three such $\bf 27$. The $E_6$ gauge symmetry can be broken as 
follows \cite{Group,Hewett:1988xc},
\begin{eqnarray}
E_6 \to\ SO(10) \times \ U(1)_{\psi} \to\ SU(5) \times\ U(1)_{\chi} \times\
U(1)_{\psi}~.~\,
\end{eqnarray}
The $U(1)_{\psi}$ and $U(1)_{\chi}$ charges for the $E_6$ fundamental ${\bf 27}$ representation 
are also  given in Table \ref{E6charge}. 

The $U(1)'$ is one linear combination of
the $U(1)_{\chi}$ and $U(1)_{\psi}$
\begin{eqnarray}
Q^{\prime} &=& \cos\theta \ Q_{\chi} + \sin\theta \ Q_{\psi}~.~\,
\label{E6MIX}
\end{eqnarray}

The other $U(1)$ gauge symmetry from the orthogonal linear combination  as well as the $SU(5)$ is  
broken at a high scale. This will allow us to have a large doublet-triplet splitting scale, which prevents rapid 
proton decay if the $E_6$ Yukawa relations were enforced. 
This will need either two pairs of (${\bf 27}$, ${\bf {\overline{27}}}$)
and one pair of (${\bf 351'}$, ${\bf \overline{351'}}$) dimensional Higgs representations,
or one pair of (${\bf 27}$, ${\bf {\overline{27}}}$), ${\bf 78}$,
and one pair of (${\bf 351'}$, ${\bf \overline{351'}}$) dimensional Higgs representations
(Detailed studies of $E_6$ theories with broken Yukawa relations can be 
found in~\cite{King:2005jy,Babu:2015psa}.) For our model, the unbroken symmetry at the TeV scale 
is $SU(3)_C \times SU(2)_L \times U(1)_Y \times U(1)'$.

\begin{table}[t]
\begin{center}
\begin{tabular}{|c| c| c| c| c|}
\hline $SO(10)$ & $SU(5)$ & $2 \sqrt{10} Q_{\chi}$ & $2 \sqrt{6}
Q_{\psi}$ & $4 \sqrt{15} Q$ \\
\hline
16   &   $10~ (Q_i, U_i^c, E_i^c )$ & --1 & 1  & $1$ \\
            &   ${\bar 5}~ ( D_i^c, L_i)$  & 3  & 1  & 7         \\
            &   $1 ~(N_i^c/T)$             & --5 & 1  & $-5$         \\
\hline
       10   &   $5~(XD_i,XL_i^c/H_u)$    & 2  & --2 & $-2$         \\
            &   ${\bar 5} ~(XD_i^c, XL_i/H_d)$ & --2 &--2 & $-8$ \\
\hline
       1    &   $1~ (XN_i/S)$                  &  0 & 4 & 10 \\
\hline
\end{tabular}
\end{center}
\caption{Decomposition of the $E_6$ fundamental  ${\bf 27}$
representation under $SO(10)$, $SU(5)$, and the $U(1)_{\chi}$,
$U(1)_{\psi}$ and $U(1)'$ charges.}
\label{E6charge}
\end{table}
We explain our convention in some details as given in Table \ref{E6charge}. Our notation is similar to 
what is used in the supersymmetric case. We have denoted the SM quark doublets, right-handed 
up-type quarks, right-handed down-type quarks, lepton doublets, right-handed charged leptons,
and right-handed neutrinos as $Q_i$, $U_i^c$, $D_i^c$, $L_i$, $E_i^c$,
and $N_i^c$, respectively. Second, in our model, we introduce three fermionic ${\bf 27}$s,
one scalar Higgs doublet field $H_u$ from the doublet of ${\bf {5}}$ of $SU(5)$,
one scalar Higgs doublet field $H_d$ from the doublet of ${\bf {\bar 5}}$ of $SU(5)$,
one scalar SM singlet Higgs field $T$ from the singlet of ${\bf 16}$ of $SO(10)$,
and one scalar SM singlet Higgs field $S$ from the singlet of ${\bf 27}$ of $E_6$. 
Thus, similar to the fermions, all the scalars with mass in the TeV scale are coming from the 
${\bf 27}$ of $E_6$. Note that the additional fermions from the ${\bf 27}$ with masses at the TeV 
scale are
 $N_i^c$, $XD_i$, $XL_i^c$, $XD_i^c$, $XL_i$, and $XN_i$. For details, please see 
 Table \ref{Particle-Spectrum}.

In our model, the $S$ gives the Majorana mass to the right-handed neutrinos $N^c_i$ after $U(1)'$
gauge symmetry breaking, {\it i.e.}, the terms $S N_i^c N_i^c$ are
$U(1)'$ gauge invariant. The mixing angle  in our model is given by 
\begin{eqnarray}
\tan\theta = \sqrt{5/3}~.~\,
\end{eqnarray}

\begin{table}[h]
\begin{tabular}{|c|c|c|c|c|c|}
\hline
~$Q_i$~ & ~$(\mathbf{3}, \mathbf{2}, \mathbf{1/6}, \mathbf{1})$~ &
$U_i^c$ &  ~$(\mathbf{\overline{3}}, \mathbf{1}, \mathbf{-2/3}, \mathbf{1})$~ &
~$D_i^c$~ & ~$(\mathbf{\overline{3}}, \mathbf{1}, \mathbf{1/3}, \mathbf{7})$ ~\\
\hline
~$L_i$~ & ~$(\mathbf{1}, \mathbf{2},  \mathbf{-1/2}, \mathbf{7})$~ &
$E_i^c$ &  $(\mathbf{1}, \mathbf{1},  \mathbf{1}, \mathbf{1})$ &
~$N_i^c/T$~ &  $(\mathbf{1}, \mathbf{1},  \mathbf{0}, \mathbf{-5})$~ \\
\hline
~$XD_i$~ & ~$(\mathbf{3}, \mathbf{1}, \mathbf{-1/3}, \mathbf{-2})$~ &
~$XL^c_i,~H_u$~ & ~$(\mathbf{1}, \mathbf{2},  \mathbf{1/2}, \mathbf{-2})$~ &
~$XD_i^c$~ & ~$(\mathbf{\overline{3}}, \mathbf{1}, \mathbf{1/3}, \mathbf{-8})$ ~\\
\hline
~$XL_i,~H_d$~ & ~$(\mathbf{1}, \mathbf{2},  \mathbf{-1/2}, \mathbf{-8})$~ &
~$XN_i,~S$~ &  $(\mathbf{1}, \mathbf{1},  \mathbf{0}, \mathbf{10})$~ &
&     \\
\hline
\end{tabular}
\caption{The particles and their quantum numbers under the
  $SU(3)_C \times SU(2)_L \times U(1)_Y \times U(1)'$ gauge symmetry. Here,
  the correct $U(1)'$ charges are the $U(1)'$ charges in the Table divided
  by $4{\sqrt{15}}$.}
\label{Particle-Spectrum}
\end{table}

The Higgs potential needed for our purpose giving rise to the extra $U(1)$ symmetry breaking is 
\begin{eqnarray}
  V = -m_S^2 |S|^2 -m_T^2 |T|^2  + \lambda_S |S|^4 + \lambda_T |T|^4
  + \lambda_{ST} |S|^2|T|^2 + (\sigma S T^2 +H.C.) ~.~\,
\end{eqnarray}
Note that without the term $\sigma S T^2$, there are two global $U(1)$ symmetries for the 
complex phases of $S$ and $T$. After $S$ and $T$ obtain the Vacuum Expectation Values (VEVs), 
we have two Goldstone bosons, and one of them is eaten by the extra $U(1)$ gauge boson. Thus, to 
avoid the extra Goldstone boson, we need the term $\sigma S T^2$ to break one global symmetry. 
Then we are left with only one global symmetry in the above potential, which is the extra $U(1)'$ gauge 
symmetry. Thus, after $S$ and $T$ acquire the VEVs, the $U(1)'$ gauge symmetry is broken, and 
$S$ and $T$ will be mixed via the $\lambda_{ST} |S|^2|T|^2$ and $\sigma S T^2$ terms.

The Yukawa couplings in our models are 
\begin{eqnarray}
  -{\cal L} &=& y_{ij}^U Q_i U_j^c H_u + y_{ij}^D Q_i D_j^c H_d + y_{ij}^E L_i E_j^c H_d
  + y_{ij}^N L_i N_j^c H_u + y_{ij}^{XNd} XL_i^c XN_j H_d \nonumber\\&&
  + y_{ij}^{XNu} XL_i XN_j H_u 
  + y_{ij}^{TD} D_i^c XD_j {T} + y_{ij}^{TL} XL_i^c L_j {T}
\nonumber\\&&
+ y_{ij}^{SD} XD_i^c XD_j S
  + y_{ij}^{SL} XL_i^c XL_j S + {\rm H. C.}~,~\,    
\end{eqnarray}
where $i=1,~2,~3$.
Thus,  after $S$ and $T$ obtain VEVs or after $U(1)'$ gauge symmetry breaking,
$(XD_i^c,~XD_i)$ and $(XL_i^c,~XL_i)$ will become vector-like particles
from the $y_{ij}^{SD} XD_i^c XD_j S$ and $y_{ij}^{SL} XL_i^c XL_j S$ terms,
and $(D_i^c,~XD_i)$ and $(XL_i^c,~L_i)$ will obtain vector-like masses
from the $ y_{ij}^{TD} D_i^c XD_j {T}$ and $y_{ij}^{TL} XL_i^c L_j {T}$ terms.
For simplicity, we assume $y_{ij}^{SD} \langle S \rangle >>  y_{ij}^{TD} \langle  {T}\rangle$
and $y_{ij}^{SL} \langle S \rangle >>  y_{ij}^{TL} \langle {T}\rangle$.
After we diagonalize their mass matrices, we obtain the mixings
between $XD_i^c$ and $D_i^c$, and the mixings between $XL_i$ and $L_i$.
The discussion of the Higgs potential for electroweak symmetry breaking is similar to
the Type II two Higgs doublet model, so we will not repeat it here.

After $U(1)'$ gauge symmetry breaking, we obtain two CP-even Higgs fields $s_h$ and $t_h$,
and one CP-odd Higgs field $a_h$ by diagonalizing the mass matrices of $S$ and $T$.
Thus, we can have one, two, or three 750 GeV resonances of which the most likely candidate would be the 
$s_h$ as it will have the strongest coupling to the vector-like fermions. 
Although it looks equally probable to have a mixture of any of the aforementioned fields to be the 
observed resonance in the diphoton mode, there is also the possibility that one can have additional 
sources for the observed resonance. There is a viable region of the parameter space in our model 
where the pair production of the vector-like fermions, $(XD_i^c,~XD_i)$, and their subsequent decays 
to any of the scalars, $s_h$, $t_h$ or $a_h$  will give rise to the diphoton and other observable 
signals. The phenomenology of the model is discussed in the next section. We note that the
supersymmetric $E_6$ GUT has also been used in Ref.~\cite{king}, but our model is non-supersymmetric. And 
more importantly {\it we address an additional mechanism for producing the diphoton excess and associated 
signals, and therefore the subsequent predictions are entirely different from theirs}.

We note that the $U(1)'$ gauge boson couples to all the SM fields in addition to
the new matter and scalar fields. The covariant derivatives for the $SU(2)_L$ doublet and the singlet scalars are given by 
\begin{align}
\label{covariant}
\begin{split}
 {\mathcal{D}}_{\mu} &= (\partial_\mu -i \frac{\vec{\sigma}}{2}.\vec{W_\mu} - i g^\prime Y  B_\mu - i g_{X} Y_{X} \Zp{_\mu}),
\end{split}
\end{align}
where $Y(H_u)=\frac{1}{2}, Y(H_d)=-\frac{1}{2}$ and $Y_X(H_u)=-\frac{2}{4\sqrt{15}}, Y_X(H_d)=-\frac{8}{4\sqrt{15}}$; 
\begin{align}
\label{covariant}
\begin{split}
 {\mathcal{D}}_{\mu} &= (\partial_\mu  - i g_{X} Y_{X} \Zp{_\mu}),
\end{split}
\end{align}
where $Y_X(S)=\frac{10}{4\sqrt{15}}$ and $Y_X(T)=-\frac{5}{4\sqrt{15}}$. 
The mass square matrix for the neutral gauge boson sector in the $(W_3,B,Z^\prime)$ basis is then 
given as
$$ \mathcal{M}=
\begin{pmatrix}
 {\Huge({\mathcal{M}_{SM}})_{2\times2}} & \begin{matrix} \mathcal{M}_{13} \\ \mathcal{M}_{23} \end{matrix} \\
\begin{matrix} \mathcal{M}_{13} & \mathcal{M}_{23}\end{matrix} & \mathcal{M} _{33},
\end{pmatrix}$$
where 
\begin{align*}
  \mathcal{M}_{13}=\frac{gg_X}{8\sqrt{15}}(2v_u^2-8v_d^2) &,&
  \mathcal{M}_{23}=-\frac{g^\prime g_X}{8\sqrt{15}}(2v_u^2-8v_d^2)~, \\
{\rm and} && \mathcal{M}_{33} = \frac{g_X^2}{240}(4v_u^2+64v_d^2+25v_t^2+100v_s^2)~,
\end{align*}
 where the $v_is$ represent
the VEVs of the scalar multiplets.
Thus, we can clearly see that the new gauge boson mass is dependent on the VEVs of all the scalars, 
such that one can choose one singlet VEV to be much smaller than the other and still have a very 
heavy $Z'$ that evades the existing limits. Moreover, the mixings between $W_3/B$ and $Z^\prime$
will be zero at tree level if $v_u=2v_d$.

The VEVs of the $SU(2)$ doublet scalars determine the SM gauge boson masses and therefore $v_{EW}=\sqrt{v_d^2 + v_u^2} \simeq 246$ GeV. 
The structure for the VEVs is given as 
\begin{align*}
  <H_d> &= \begin{pmatrix} v_d/\sqrt{2} \\ 0 \end{pmatrix}~,~ & <H_u> &= \begin{pmatrix} 0 \\ v_u/\sqrt{2}\end{pmatrix}~,~ \\
  <T> &= v_t/\sqrt{2}~,~ & <S> &= v_s/\sqrt{2}~,~
\end{align*}
which leads to the following mass matrix for the down-type quarks and the charged leptons in the 
$(q_1,q_2,q_3,xq_1,xq_2,xq_3)$ basis is given as
\begin{align}
  \frac{1}{\sqrt{2}}
\begin{pmatrix}
y_{ij}^D v_d &  0\\
y_{ji}^{TD}v_t  & y_{ij}^{SD}v_s 
\end{pmatrix},  &&
  \frac{1}{\sqrt{2}} 
\begin{pmatrix}
y_{ij}^E v_d &  y_{ji}^{TL}v_t  \\
 0 & y_{ij}^{SL} v_s 
\end{pmatrix},
\end{align}
where $i,j=1,2,3$. The $q_i$s and $xq_i$s represent the down-type quarks for the left matrix and the 
leptons for the right matrix. These mass matrices would be diagonalized by a bi-unitary 
transformation which would 
lead to a mixing between the vector-like fermions and the SM fermions. However, one should note that 
the mixings between the left-handed fermions and the right-handed fermions will be dictated by a different
set of mixing angles. This would have significant implications in the rest of our analysis and plays a crucial
role in the signal we have proposed.   

We should also point out a few useful assumptions that we think are relevant for the analysis:

\begin{enumerate}
  \item We have neglected any mixing between the electroweak doublet scalars and singlet scalars. 
  \item We also ensure that the new $U(1)'$ gauge boson does not have a significant mixing with the 
  SM gauge boson $Z$ ($\mathcal{M}_{13}, \mathcal{M}_{23} << \mathcal{M}_{33}$). 
  \item The mixings between the left-handed SM fermions and the vector-like fermions are taken to be zero, {\it i.e.},
  we assume the zero left-mixing angle ($\theta_L=0$). This will insure that the vector-like fermions do not 
  decay to the SM gauge bosons and light SM fermions~\cite{Grossmann:2010wm}, thus allowing the only 
  significant channel that would be the decay to a scalar ($s_h, t_h$) and SM fermion. One can also get a 
  very suppressed partial width (due to the smallness of the Yuakawa couplings courtesy the masses 
  of the SM down-type quarks and leptons) that decays to a SM Higgs and light SM fermion, through 
  the mixings in the right-handed fermion sector.
\end{enumerate}

\section{Analysis}\label{sec:analysis}
The diphoton signal in our model can arise from several possibilities of the particle spectrum as 
well as in two different production channels. Note that the most popular option in the literature in 
explaining the diphoton excess has been through the on-shell production of a 750 GeV resonance via
gluon fusion which then decays to diphotons. Therefore we must point out that such a possibility 
clearly exists in our model description. We shall show that what part of the parameter space 
is best suited for the aforementioned explanation of the diphoton excess and what sort of masses and 
couplings are required for the new exotics to satisfy the experimental data. Notwithstanding this possibility, 
we wish to also highlight a new channel of production that can also give rise to the diphoton excess 
signal and predict an accompanying dijet resonance at the same mass in the same signal events. 
This would be possible through the pair production of vector-like quarks which is heavier but very close 
in mass to one of the scalars which has mass of 750 GeV. The pair produced exotic quarks then decay 
to this 750 GeV scalar and a very soft jet, which is not detected. So we get a pair of these 750 GeV 
scalars which then decay to either a gluon pair or a diphoton pair. We believe that such a possibility, which
exists for a range of the VLQ mass that nearly degenerates with the resonant scalar mass and 
yet is searched for by the LHC experiments, should show itself with a closer scrutiny of the events along 
with the new events being collected.  We now take up these two possible channels and analyze the 
diphoton signal for different parameter choices of the model. To calculate the several branching fractions 
of the new exotic particles and signal cross sections we have used the {\tt CalcHEP} event 
generator \cite{Belyaev:2012qa}. We have implemented the model in {\tt LanHEP} \cite{lanhep} 
to create the model files for {\tt CalcHEP}.

Note that the analysis will vary depending on the number of vector-like fermions we might possibly have
at sub-TeV masses. The colored vector-like fermions participate in both the production channels while the
non-colored one's help in increasing the diphoton branching fraction of the scalar. 
We shall consider three different scenarios:
\begin{itemize}
 \item Only one vector-like quark with mass around a TeV while all the remaining vector-like fermions 
 have mass of 1.5 TeV.  
 \item One vector-like quark and one vector-like lepton having TeV mass while the remaining have mass 
 of about 1.5 TeV.
 \item One vector-like quark and two vector-like degenerate leptons having TeV masses while again the rest 
 of them have mass at 1.5 TeV.
\end{itemize}

Note that the choice of a fixed scale of 1.5 TeV mass for the vector-like fermions is made as it
helps in expressing our results in terms of the singlet scalar VEV of $S$ given by $v_s$ which 
gives the mass to the vector-like fermions as $M_{xd_{i}} = y_{ii}^{SD}v_s/\sqrt{2}$ and 
$M_{x\ell_{i}} = y_{ii}^{SL}v_s/\sqrt{2}$. In addition, it helps us in scanning a range for $v_s$ such 
that the perturbative limit of the Yukawa couplings  (taken as $y_{ii}/\sqrt{2}<\sqrt{4\pi}$) is not violated.
Therefore, a shift in the choice of the mass of the heaviest vector-like fermion would imply that $v_s$ 
can be varied over a different range. This gives us the necessary dynamics to understand how the signal 
strength gets affected by the choice of the mass of the vector-like fermion.

To study the final states we specify here the parameters that are relevant for the analysis. The vector-like 
quarks and leptons are represented as $xd_{i}$ and $x\ell_{i}$ respectively. As pointed out 
earlier, the VEVs for $S$ and $T$ which are given by $v_s$ and $v_t$ respectively 
play a significant role in giving mass to the $U(1)^\prime$ gauge boson, which we call  $Z^\prime$. We 
take the mass of $Z^\prime$ to be 1.5 TeV, and fix the value of $v_t$ to be 10 TeV. Such large value of 
$v_t$ becomes a necessity to avoid a significant mixing in the neutral gauge boson sector. This 
also forces very small mixing between the scalars $s_h$ and $t_h$. This decoupled 
particle spectrum is somewhat also preferable as the LHC has not observed any other signal other than the 
diphoton resonance. Note that the new particles in the spectrum can however play a crucial role in 
determining additional signals for the model at the LHC, which we leave for future analysis as we 
focus only on the diphoton signal in this work. We consider two types of processes that will contribute to the 
diphoton production cross section. Note that the experiments have observed a resonance at 750 GeV in 
the diphoton channel which is best explained by the direct on-shell production in the $s$-channel as 
shown in Fig.~\ref{fig:feynman1}. For simplicity, we will take all types of  Yukawa couplings $y^A_{ij}$   
to be zero for $i\neq j$, where $A\equiv TD, TL, SD, SL$ (see Sec. \ref{sec:model}). 

\begin{itemize}  
 \item $pp\rightarrow s_h \rightarrow \gamma\gamma$
 
\begin{figure}[t!]
\includegraphics[width=10cm]{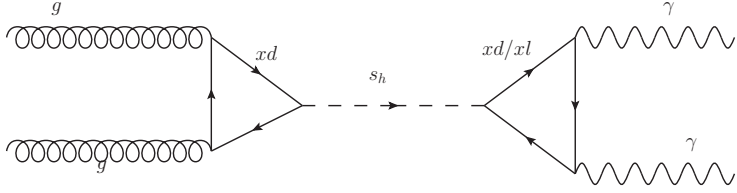}
\caption{The Feynman diagram which contributes to the diphoton production through the onshell production 
of $s_h$ as an s-channel resonance via gluon-gluon fusion at the LHC.}
\label{fig:feynman1}
\end{figure}
 Through out the analysis we shall identify this process as diphoton production through resonant 
 $s_h$ channel. The parton level Feynman diagram for this process is shown in Fig. \ref{fig:feynman1} 
 and the cross section is given by 
 $\sigma_1 = \sigma(pp\rightarrow s_h)\times Br(s_h\rightarrow \gamma\gamma)$. 
\begin{figure}[h!]
\includegraphics[width=.60\linewidth,height=3.1in]{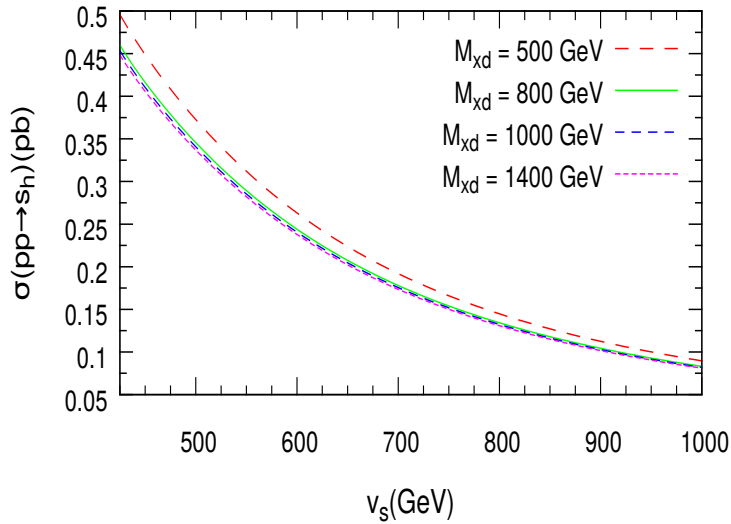}
\caption{The s-channel production cross section of $s_h$ via the gluon-gluon fusion at LHC 
with $\sqrt{s}=13$ TeV, as a function of VEV of the singlet $S$; $v_s$ for different values of 
$M_{xd_{1}}$.}
 \label{fig:sh_production}
\end{figure}
To study the $s_h$ production, the relevant parameters are the VLQ mass and the VEV $v_s$ which 
gives the 
Yukawa strength of the VLQ interaction with the scalar $s_h$.
The loop induced production of the $s_h$ depends on its coupling to the massless gluons 
which is given by the effective Lagrangian
\begin{align} \label{eq:sgg}
\mathcal{L}_{sGG} = - \lambda_{sgg} s ~G_{\mu\nu} G^{\mu\nu}. 
\end{align}
In Eq. \ref{eq:sgg}, $\lambda_{sgg} = \alpha_s  F_{1/2} (\tau_{xd}) /( 16 \pi v_s) $ where 
$F_{1/2}(\tau_{xd}) = 2(\tau_{xd}+(\tau_{xd}-1)f(\tau_{xd}))\tau_{xd}^{-2}$ represents the loop function and 
$f(\tau_{xd})=(sin^{-1}\sqrt{\tau_{xd}})^2$ with $\tau_{xd}=m_{s_h}^2/4M_{xd}^2$. 
The $s_h$ production cross section depends on the VLQ content 
only and not on the VLL content. For all the three cases we have considered for our analysis, we choose 
only one VLQ whose mass is varied while the remaining two have mass of 1.5 TeV. 
Hence for all the three cases the $s_h$ production cross section will be same for a given value 
of  $M_{xd_{1}}$. In Fig. \ref{fig:sh_production} we show the leading-order production 
cross section of the scalar $s_h$ as a function of the VEV $v_s$ for different values of the 
lightest VLQ mass. The lower values of $v_s$ imply a large Yukawa coupling invariably leading to larger
production cross sections. The cross sections are doubled if one takes a QCD $K$-factor of $\sim 2$ 
as for the SM Higgs boson. In addition, we note that as the lightest VLQ mass is increased to 1.4 TeV,
there does not seem to be a significant fall. This is because in all production curves, there is some
contributions from the two heavy VLQ's with mass 1.5 TeV. Thus, with even three VLQ's with mass 1.5 TeV,
the production cross section can be quite large.

 \item $pp\rightarrow xd$  $\overline{xd}\rightarrow \gamma\gamma$ + jets 
\begin{figure}[h!]
\includegraphics[width=7cm]{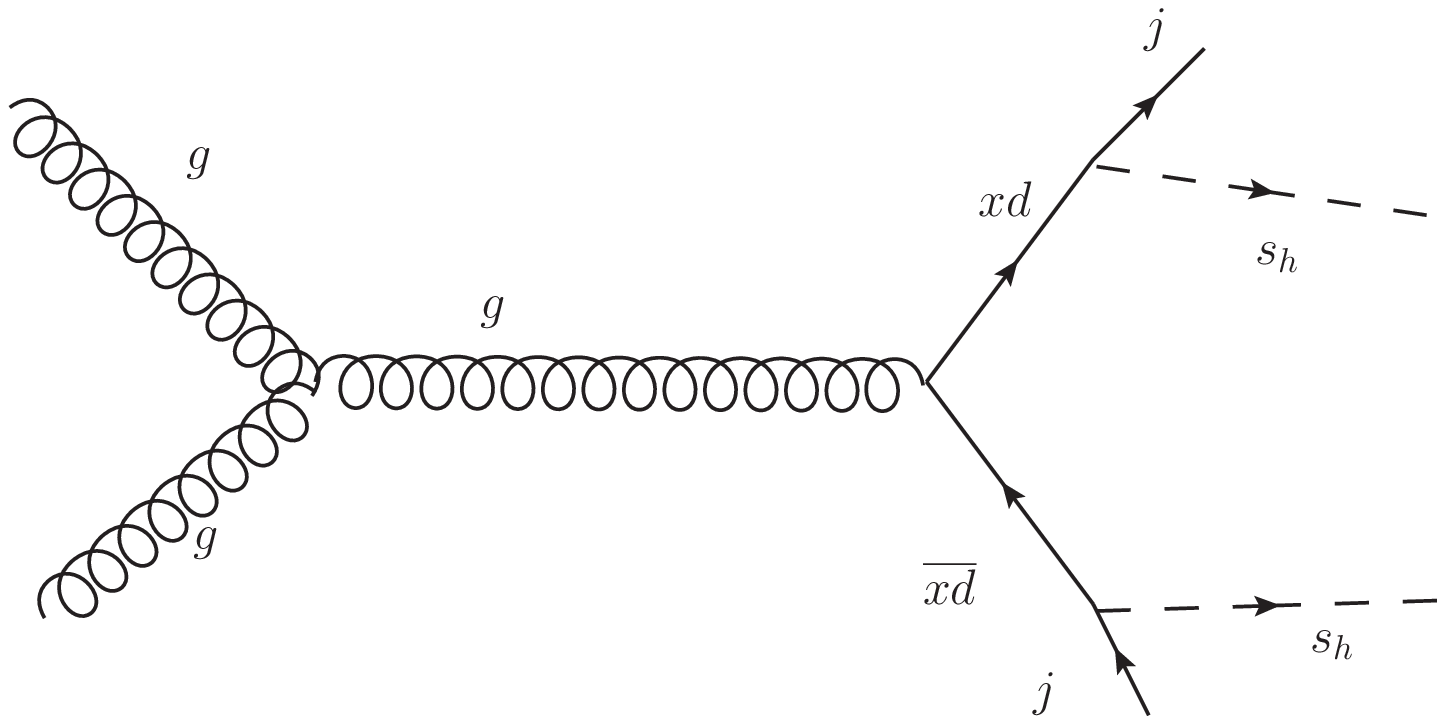}
\includegraphics[width=7cm]{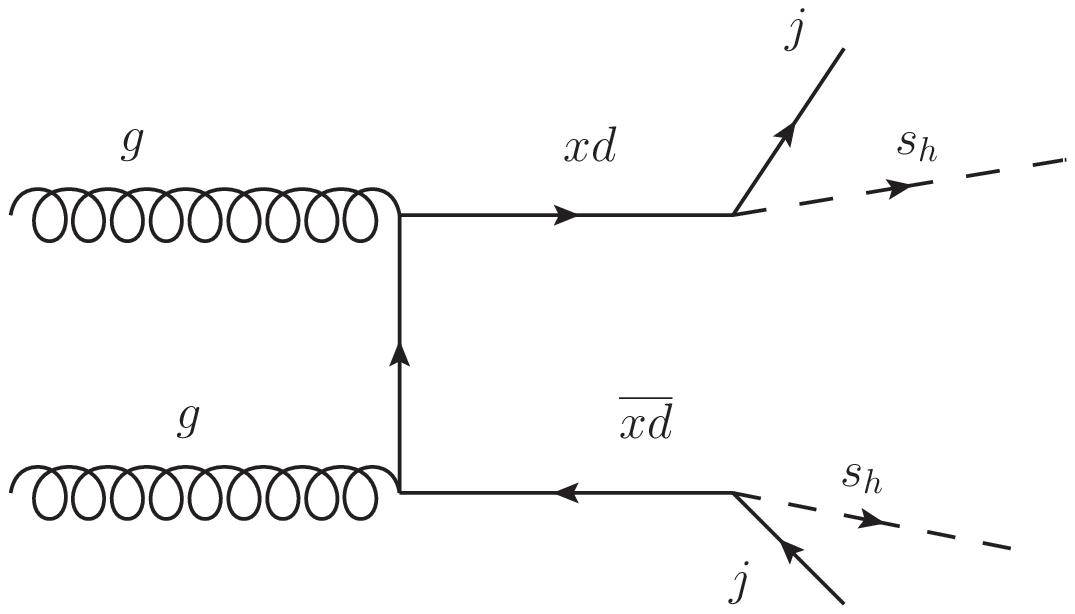}
\caption{The Feynman diagrams for the dominant subprocess contributing to the pair production of the VLQ and 
its subsequent decay to $s_h$ and an additional jet ($j$) giving an $s_h$ pair and two jets in the final state.}
\label{fig:feynman}
\end{figure} 

This is the alternative approach that can also contribute to the events of diphoton excess. 
Through out the analysis this process is termed as diphoton production through $xd$ pair production channel. 
The parton level Feynman diagrams for the dominant process are shown in Fig.~\ref{fig:feynman}, where we have
not shown the diagrams for the $q\bar{q}$ initiated subprocesses. 

Note that the production of the $xd$ pair at LHC is dominantly through the gluon-gluon fusion 
and in Fig. \ref{fig:xd_production} we plot the total pair production cross section (leading-order) 
of $xd$ at the LHC  with $\sqrt{s}=13$ TeV, as a function of $M_{xd}$ with values between 760-800 
GeV. The decay $xd\rightarrow s_h \,\, j$ leads to a relatvely soft jet and a pair of the 750 GeV scalar $s_h$.  
The $s_h$ can now decay to a pair of photons or gluons. Thus we get a resonant diphoton signal along 
with multiple jets.
\begin{figure}[h!]
\includegraphics[width=.60\linewidth,height=3.1in]{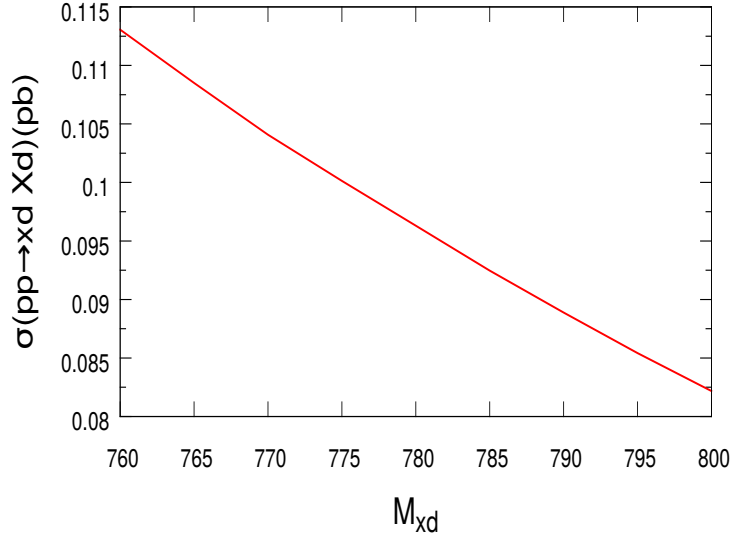}
\caption{The pair production cross section of $xd$ at LHC 
with $\sqrt{s}=13$ TeV, as a function of $M_{xd}$ with values between 760-800 GeV.}
 \label{fig:xd_production}
\end{figure}
The diphoton cross section in this channel is then given by 
 $$\sigma_2 = 2\times\sigma(pp\rightarrow xd,\overline{xd})\times [Br(xd\rightarrow s_h \,\, j)]^2 \times 
  Br(s_h\rightarrow\gamma\gamma)\times Br(s_h \rightarrow gg) .$$
For our analysis we choose parameters such that $Br(xd\rightarrow s_h \,\, j) \simeq 1$ is ensured 
when $M_{xd} > m_{s_h}$.
We have already discussed in Sec. \ref{sec:model} why the other usual decay modes of the VLQ 
are absent in our model. 
This also ensures that the mass bounds on the VLQ can be significantly relaxed as the dominant 
decay of the VLQ when $M_{xd} < m_{s_h}$ is $xd \rightarrow h \,\, j$ while $xd \rightarrow s_h \,\, j$ 
when $M_{xd} > m_{s_h}$. Note that the decay to the SM gauge bosons will be allowed once we 
allow the left-handed SM fermions to mix with the VLQ and VLL.
\end{itemize}

\begin{figure}[t!]
\includegraphics[width=.60\linewidth,height=3.1in]{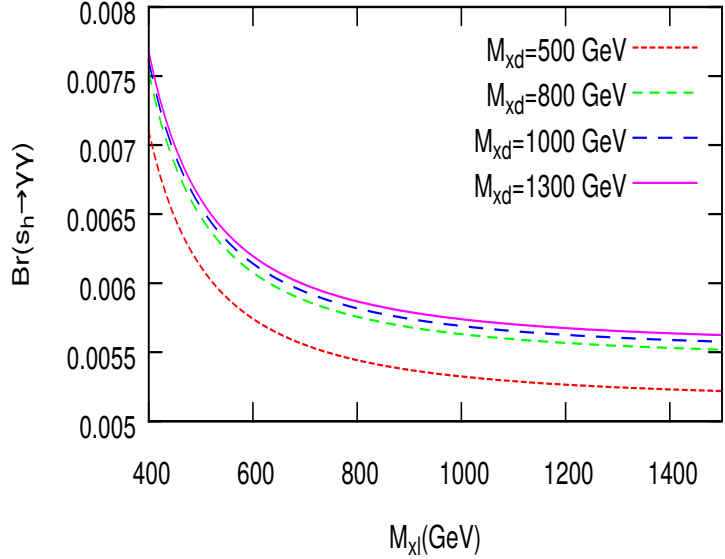}
\caption{The diphoton branching ratio as a function of the VLL mass for different values of the lightest VLQ
mass. The branching ratio is evaluated using two light degenerate VLL's with mass $M_{x\ell}$ and one 
light VLQ with mass $M_{xd}$.}
 \label{fig:braa}
\end{figure}

The decay of the scalar into two photons is again dependent on the mass and Yukawa strengths 
of the vector-like fermions. As the number of light vector-like leptons are not severely constrained 
by experiments and their inclusion helps in improving the $s_h \to \gamma \gamma$ branching fraction,
we shall consider the results with one or more of such VLL contributing to the diphoton decay. We 
plot the branching ratio for the scalar $s_h$ decaying into a pair of photons in Fig. \ref{fig:braa} 
as a function of the VLL mass for different values of the lightest VLQ mass. For the branching ratio plot, we 
have considered two degenerate VLL's  and one light VLQ while the heaviest VLL and VLQ's have 
their mass set 
at 1.5 TeV. Note that including more light VLQ's would boost the production channel for both the 
production processes ($pp \to s_h$ and $pp \to xd \,\, \overline{xd}$) but reduces the branching fraction of 
$s_h \to \gamma \gamma$ as the partial width of $s_h \to gg$ becomes much larger. We therefore now 
consider the two cases where we have one light VLQ and VLL each, and when we have one light VLQ and 
two light degenerate VLL's.  

\subsection{Diphoton signal}

We now proceed to analyse the two production channels for the scalar $s_h$ and compare the signal 
strengths into the diphoton mode.  
Being in the perturbative limit of the Yukawa couplings and by varying parameters $v_s$, $M_{xd}$ 
and $M_{x\ell}$ we shall check the availability of parameter space which keeps the diphoton cross section 
within 3-10 fb range. The light vector-like leptons have been taken to be degenerate for simplicity. 
The value of $v_s$ has been varied from 425 GeV to 1 TeV with the lower cut-off, dependent on the 
choice of 1.5 TeV as the mass of the heaviest VLQ/VLL which determines the perturbative limit for 
the Yukawa couplings. Thus, it is clear that if the mass of the heaviest VLQ is reduced, then a lower $v_s$ 
would be allowed. The $s_h$ production cross section decreases with increase in either $M_{xd}$ or 
$v_s$ as shown in Fig.\ref{fig:sh_production}. This happens because the loop contributions 
get suppressed for heavier $M_{xd}$ while the Yukawa strength becomes smaller when $v_s$ is 
increased for a fixed $M_{xd}$.

\begin{figure}[t!]
 \includegraphics[width=3.0in,height=2.6in]{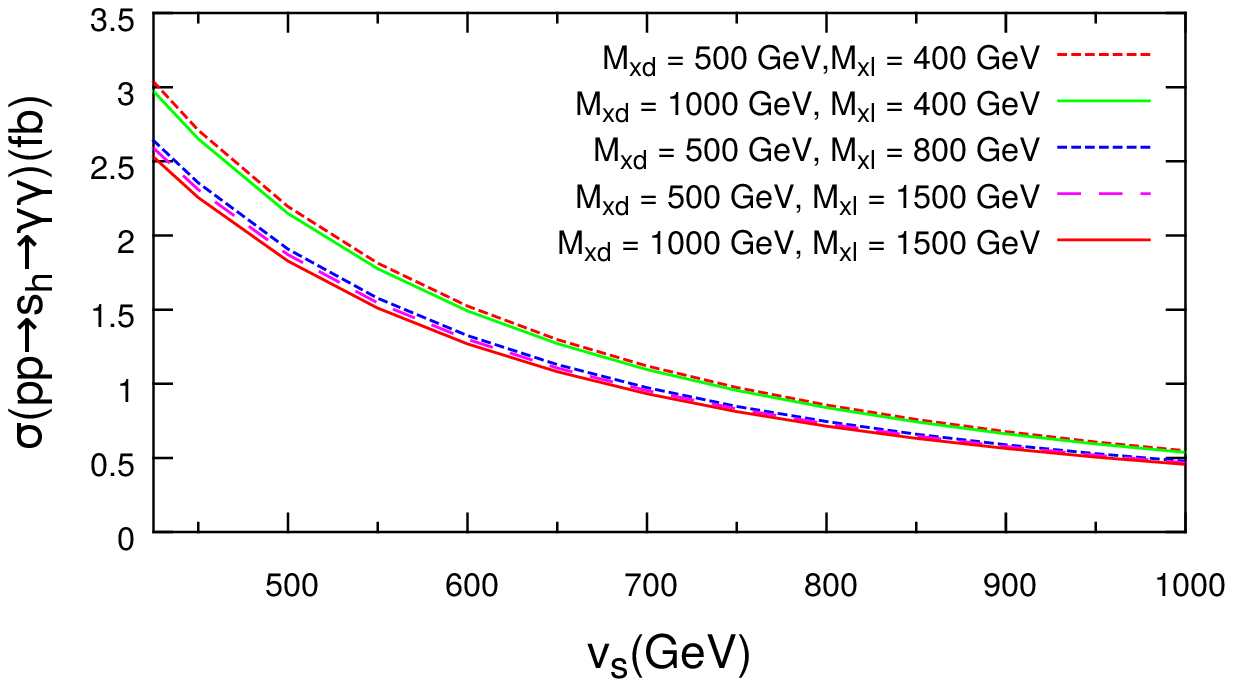}
  \includegraphics[width=3.0in,height=2.6in]{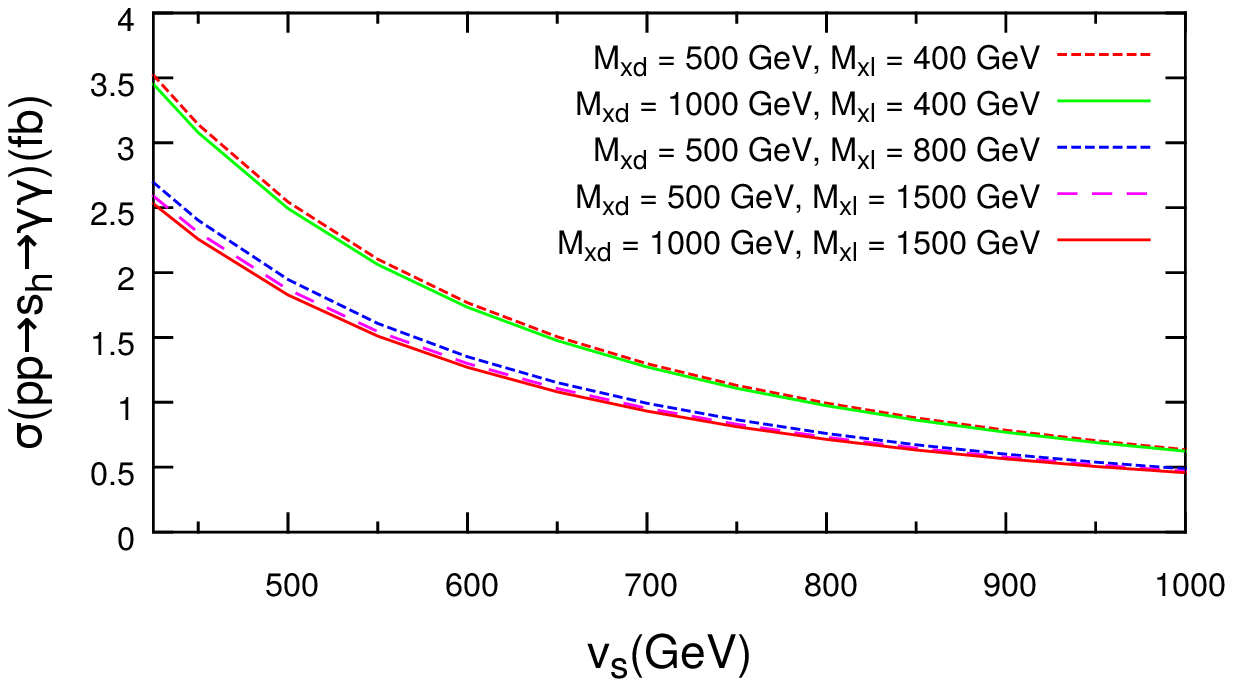}
 \caption{The diphoton production cross section as a function of $v_s$ for different choices of 
 $M_{xd}$ and $M_{x\ell}$. The left panel represents (one VLQ, one VLL) while the right-panel represents 
 (one VLQ, two VLL) whose mass is varied. The heavier VLQ/VLL states have a fixed mass of 1.5 TeV.}
 \label{fig:sh_aa_production}
 \end{figure}
In Fig.~\ref{fig:sh_aa_production} we show the contribution to diphoton cross section from $s_h$ 
resonant production channel as a function of $v_s$ for different sets of ($M_{xd}$,$M_{xl}$). Note that 
the left panel in Fig. \ref{fig:sh_aa_production} corresponds to the case of (one VLQ, one VLL) while the
right panel represents (one VLQ, two VLL) whose masses are varied. The curves also represent a 
variation of the Yukawa coupling which is decreasing from left-to-right and quite clearly, a large value 
of the coupling is more favorable to generate the correct size of the diphoton signal rate but looks 
possible even with a VLQ as heavy as 1 TeV and a single light VLL in the spectrum. Note that a reasonable
enhancement is possible with appropriate $K$-factors for the production of $s_h$. The inclusion of 
an extra light VLL in the spectrum of similar mass seems to enhance the diphoton rate by about 16\% 
when the VLL's are in the mass range of 400 GeV, as shown in the right panel of Fig. \ref{fig:sh_aa_production}. For the heavier VLL's the difference is hardly noticeable when compared 
to the case with one light VLL. 

\begin{figure}[h!]
   \includegraphics[width=3.0in,height=2.6in]{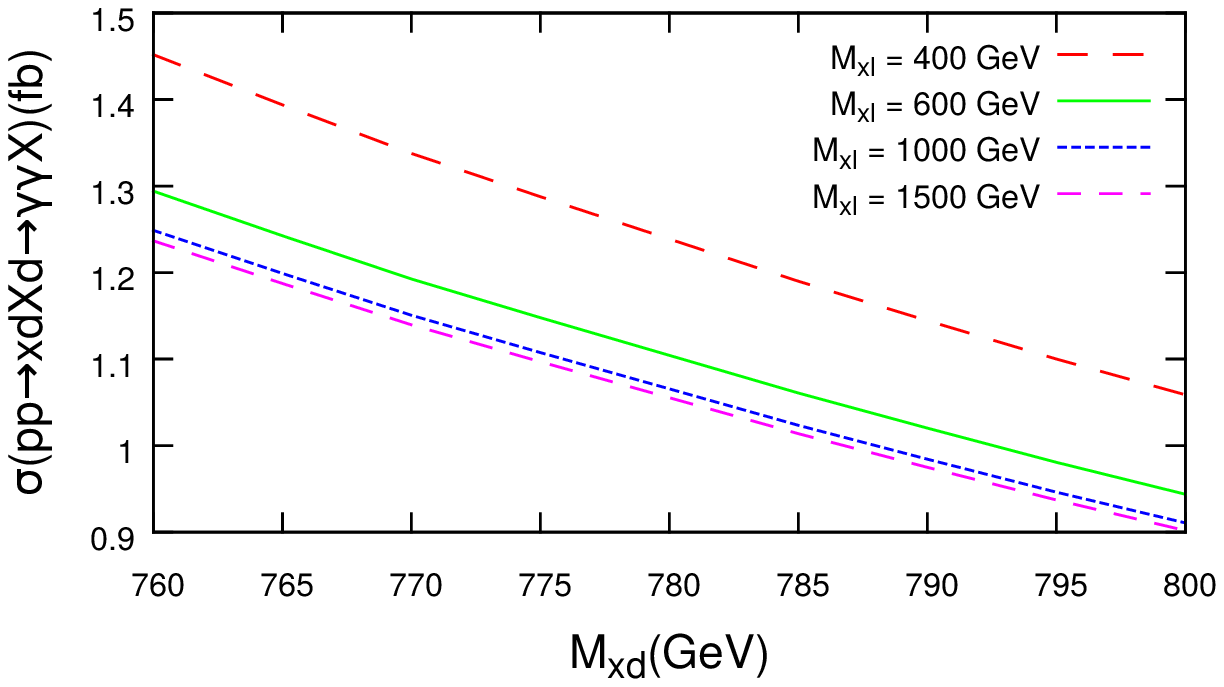}
      \includegraphics[width=3.0in,height=2.6in]{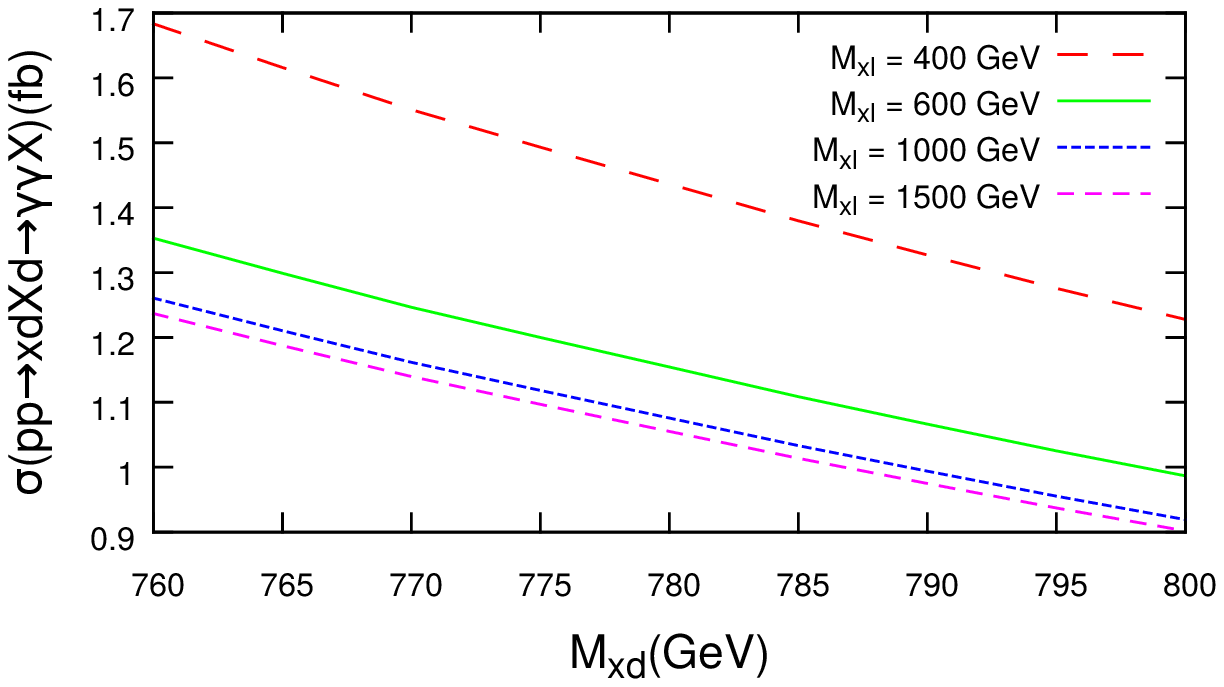}
 \caption{The inclusive diphoton production cross section from $xd$ pair production as a function of 
 $M_{xd}$ for different values of $M_{xl}$. The two panels represent the similar configurations as in 
   Fig.~\ref{fig:sh_aa_production}.}
 \label{fig:diphoton_xd}
 \end{figure}

We now consider the diphoton rate expected from the pair production of the VLQ's having mass which is
near degenerate with the 750 GeV scalar $s_h$. We assume a range of $M_{xd}=760-800$ GeV to 
show the inclusive rate for the diphoton signal. 
In Fig.\ref{fig:diphoton_xd} we plot this contribution evaluated at leading-order, to the diphoton signal 
via $xd$ pair production as 
a function of $M_{xd}$ being in the range $760-800$ GeV and for different values of $M_{x\ell}$ 
for the two cases, (one VLQ, one VLL - left panel) and (one VLQ, two VLL - right panel). Again, we must 
point out here that a reasonable enhancement is expected from appropriate QCD $K$-factors for the 
$xd$-pair produced through strong interactions. As the production of the VLQ pair as well as its decay
are not dependent on the choice of $v_s$ or the Yukawa strength, the number of light VLQ's and light 
VLL's  are the major players in determining the event rates here. However, the important thing to note 
here is that with a very degenerate VLQ and $s_h$, the contributions to the inclusive diphoton rate is 
quite significant. In principle, inclusion of more generations of the VLQ should give a much enhanced 
rate for the inclusive diphoton production in this channel which would be clearly observable in its 
own individuality.
Thus, this channel would stand to explain the diphoton excess even when the Yukawa couplings are very small 
and contributions to the diphoton rate via $s_h$ resonant production becomes negligible. However, a 
simultaneous observation of a dijet resonance at the same mass (750 GeV) would be required to 
confirm this production mode. Notwithstanding this contribution to the observed diphoton signal, this 
channel is interesting in its own respect as a confirmatory signal of the low lying spectrum that 
gives rise to the diphoton resonance, even with the VLQ's much heavier. It also suggests an alternative 
channel for VLQ searches at the LHC which complements the observed diphoton signal.
Note that the contribution will go down with increasing $M_{xd}$ as the $xd$ pair production cross section
falls for higher $M_{xd}$. However, the increase in the number of light VLL's helps in increasing the 
diphoton branching fraction and signal rate.  As long as only a single $xd$ pair production channel is 
concerned, it is quite clear from Fig.~\ref{fig:diphoton_xd} that a very limited range of parameter space 
is allowed which satisfies the diphoton cross section to be above 3 fb for the $K$-factor equal to 2.

\begin{figure}[t]
 \includegraphics[width=2.9in,height=2.5in]{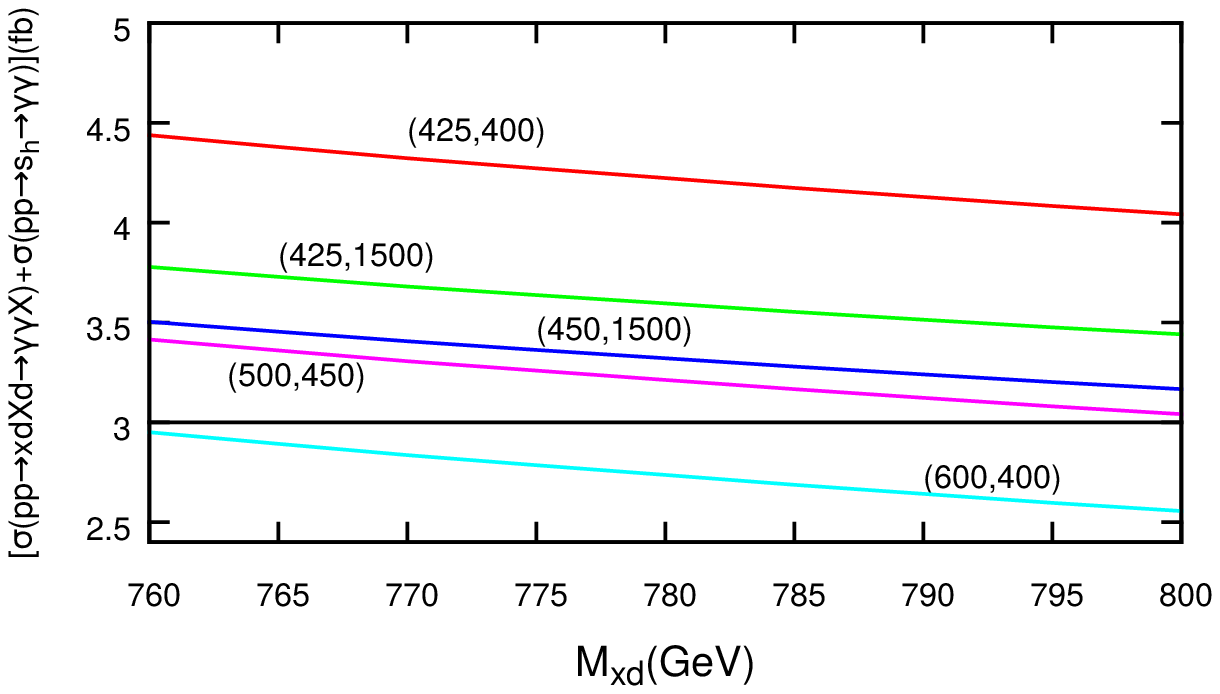}
  \includegraphics[width=2.9in,height=2.5in]{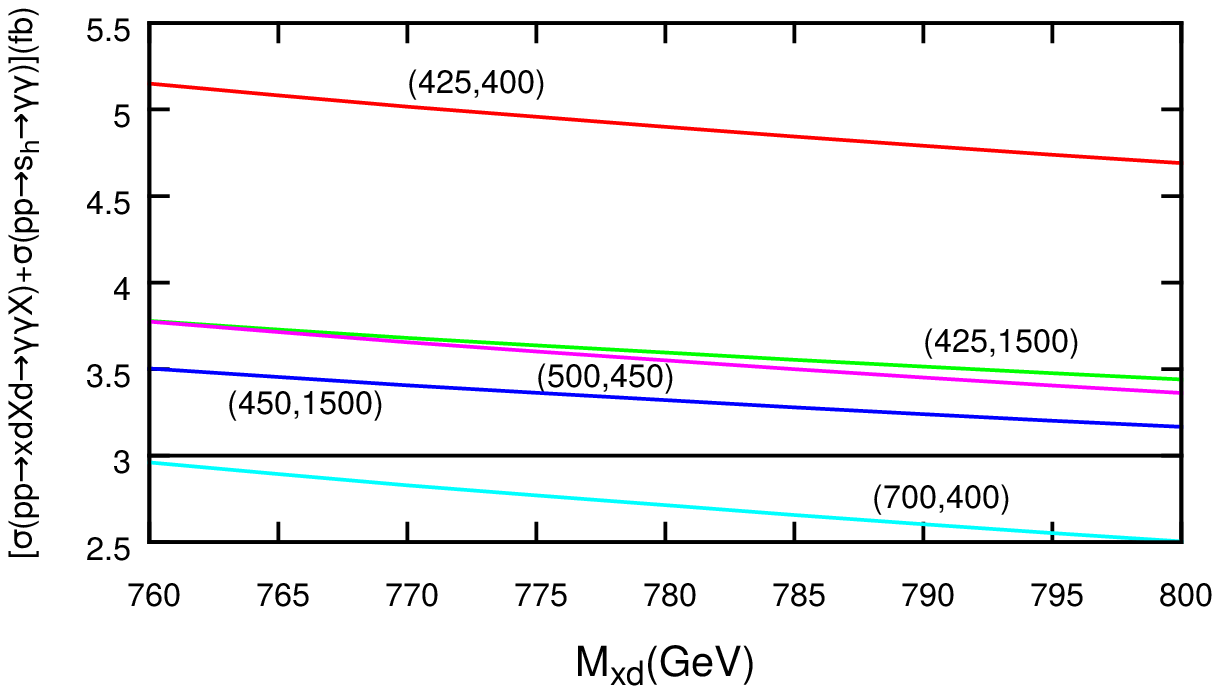}
 \caption{The combined contribution to the inclusive diphoton production cross section from 
 $xd$ pair production channel and $s_h$ resonant production channel as a function of 
  $M_{xd}$ with different choices of $(v_s,M_{xl})$. The two panels represent the same configurations 
  as in Fig.\ref{fig:diphoton_xd}.}
        \label{fig:diphoton_xd_sh}
\end{figure}

Finally we combine the contributions through both production channels for the inclusive diphoton 
cross section for the reduced mass range of the VLQ between $760-800$ GeV and plot the two 
cases of one VLL (left-panel) and two VLL's (right-panel) in Fig.~\ref{fig:diphoton_xd_sh}. The 
combination clearly shows that a significant amount of parameter space satisfying the required 
cross section opens up after taking the contribution from both the channels. The rates are 
shown in Fig. \ref{fig:diphoton_xd_sh}  from both the channels for different sets of $(v_s,M_{xl})$, with 
the fact that larger values of $v_s$ correspond to smaller values of the Yukawa couplings.
For $v_s$ value of around 700 GeV and above, it is difficult to get enough event rates at the leading order 
with the assumed values of $M_{xd}$ and $M_{xl}$ and the number of light generations. However, 
an addition of another light VLL in the range of 500 GeV would again push the cross sections up. 
Thus, it is quite possible to contemplate a wide range of values and combinations that can easily 
accommodate the diphoton excess in the current framework. 

\section{Conclusions}\label{sec:summary}
In this work we have considered an $E_6$ motivated extension of the SM where the larger symmetry 
groups are broken at a very high scale and a residual $U(1)$ gauge symmetry is the only remaining 
symmetry beyond the unbroken SM gauge symmetry. This additional $U(1)$ then gets broken at the 
TeV scale through new scalar SM singlets giving rise to a TeV scale particle spectrum with
three generations of vector-like quarks and leptons and several neutral scalars. We proposed one of the 
singlet dominated scalar to be the observed $750$ GeV resonance and that the diphoton resonance 
signal indicated by the recent LHC data might also arise from the pair production of
vector-like down-type quarks with mass a little bit heavier than $750$ GeV scalar.
The vector-like quarks decay into 
the ordinary light quark and  $750$ GeV SM singlet scalar. The subsequent decay of the scalar 
singlet produces this diphoton resonance. We also showed that there is a wide range of parameter space 
in the model that could accommodate the diphoton signal, either through the more popular 
proposals in the literatures where the scalar is produced as an $s$-channel resonance 
through gluon-gluon fusion via VLQ loop-mediated processes as well as the new channel mentioned 
above. The prediction of such a proposal in the current theoretical framework would imply an
accompanying dijet signal at the same mass with similar cross section in the $2\gamma + 2j$ final 
state in addition  to two dijet resonances at the same mass for a $4j$ final state with the cross 
sections about 100 times larger. Both the predictions would be verifiable as the luminosity
accumulates in the upcoming runs of the LHC. We also proposed that the new production channel is a new search 
mode for vector-like quarks and would severely affect the current limits on vector-like quark mass 
which rely on its decay to SM gauge bosons and quarks. Thus, even if the 750 GeV diphoton signal 
indeed proves to be a fluctuation and does not survive the scrutiny of time and the upcoming high luminosity data at the LHC, the new signal for the VLQ, proposed in this work, could provide to be an interesting channel to search for new physics beyond the SM.   

\begin{acknowledgments}
We thank K. S. Babu for useful discussions.  This research was supported in part by the Natural Science 
Foundation of China under grant numbers 11135003, 11275246, and 11475238 (TL). The work of SN was 
in part supported by US Department of Energy Grant Numbers DE-SC 001010 and DE-SC 0016013. 
The work of KD and SKR was partially supported by funding available from the Department of Atomic 
Energy, Government of India, for the Regional Centre for Accelerator-based Particle Physics (RECAPP), 
Harish-Chandra Research Institute.
\end{acknowledgments}


\end{document}